\documentclass[sigconf]{acmart}
\AtBeginDocument{%
  \providecommand\BibTeX{{%
    \normalfont B\kern-0.5em{\scshape i\kern-0.25em b}\kern-0.8em\TeX}}}

\setcopyright{iw3c2w3}

\usepackage{booktabs} 
\usepackage{tikz, pgfplots}
\pgfplotsset{compat=1.14} 
\usepackage{graphicx}
\usepackage{dblfloatfix}
\usepackage{breqn}
\usepackage{paralist}
\usepackage{balance} 
\usepackage{mathtools}
\usepackage{hyperref}
\usepackage{xcolor}
\usepackage{xspace}
\usepackage[captionskip=1pt,farskip=2pt]{subfig}
\usepackage{wrapfig}
\usepackage[justification=justified]{caption}
\usepackage{placeins}
\usepackage{array}
\usepackage{enumitem}
\usepackage{makecell}
\usepackage{booktabs}
\usepackage{multirow}
\usepackage{siunitx}

\hypersetup{
	colorlinks,
	linkcolor={red!20!black},
	citecolor={green!20!black},
	urlcolor={blue!20!black}
}
\newcommand{\doc}[1]{\ensuremath{d^{(#1)}}}

\newcommand{\trainjudg}[1]{\ensuremath{y^{(#1)}}}
\newcommand{\predictjudg}[1]{\ensuremath{\hat{y}^{(#1)}}}
\newcommand{\feat}[1]{\ensuremath{x^{(#1)}}}
\newcommand{\lfair}[0]{\ensuremath{L_{\mathit{DELTR}}}}
\newcommand{\secmargin}{\vspace{-1.5mm}}
\newcommand{\subsecmargin}{\vspace{-2.5mm}}
\newcommand{\sparamargin}{\vspace{-1mm}}

\newcommand{\scattersize}{0.33}
\newcommand{\scatterheight}{1.7in}

\newcommand{\spara}[1]{\vspace{+1mm}\noindent{\bf #1}}

\newcommand{\methodname}{\textsc{DELTR}\xspace}
\newcommand{\methodnameexpanded}{Disparate Exposure in Learning To Rank}

\newcommand{\baseline}{\textsc{FA*IR}\xspace}

\title[Disparate Exposure in Learning to Rank]{\mbox{Reducing Disparate Exposure in Ranking:} \mbox{A Learning To Rank Approach}}

\author{Meike Zehlike}
\affiliation{%
	\institution{Humboldt Universit\"at zu Berlin}
	\institution{Max-Planck-Institut for Software Systems}
}
\email{meikezehlike@mpi-sws.org}

\author{Carlos Castillo}
\affiliation{%
	\institution{Universitat Pompeu Fabra}
}
\email{chato@acm.org}

\renewcommand{\shortauthors}{Zehlike and Castillo}

\begin{abstract}
	A clear and well-documented \LaTeX\ document is presented as an
	article formatted for publication by ACM in a conference proceedings
	or journal publication. Based on the ``acmart'' document class, this
	article presents and explains many of the common variations, as well
	as many of the formatting elements an author may use in the
	preparation of the documentation of their work.
\end{abstract}

\copyrightyear{2020}
\acmYear{2020}
\acmConference[WWW '20 Companion]{Companion Proceedings of the Web Conference
	2020}{April 20--24, 2020}{Taipei, Taiwan}
\acmBooktitle{Companion Proceedings of the Web Conference 2020 (WWW '20
	Companion), April 20--24, 2020, Taipei, Taiwan}
\acmPrice{}
\acmDOI{10.1145/3366424.3383534}
\acmISBN{978-1-4503-7024-0/20/04}



\begin{document}

\linespread{0.9}

\graphicspath{{pics/}}


\begin{abstract}
Ranked search results have become the main mechanism by which we find content, products, places, and people online.
Thus their ordering contributes not only to the satisfaction of the searcher, but also to career and business opportunities, educational placement, and even social success of those being ranked.
Researchers have become increasingly concerned with systematic biases in data-driven ranking models, and various \emph{post-processing} methods have been proposed to mitigate discrimination and inequality of opportunity.
%
%
This approach, however, has the disadvantage that it still allows an unfair ranking model to be trained.
%

In this paper we explore a new \emph{in-processing} approach: \methodname, a learning-to-rank framework that addresses potential issues of \emph{discrimination} and \emph{unequal opportunity} in rankings at training time.
We measure these problems in terms of discrepancies in the \emph{average group exposure} and design a ranker that optimizes search results in terms of relevance \emph{and} in terms of reducing such discrepancies.
We perform an extensive experimental study showing that being ``colorblind'' can be among the best or the worst choices from the perspective of relevance and exposure, depending on how much and which kind of bias is present in the training set.
We show that our in-processing method performs better in terms of relevance and exposure than a pre-processing and a post-processing method across all tested scenarios.
\vspace{-1.5mm}
\end{abstract}

\maketitle
\section{Introduction}
\label{sec:introduction}

Ranked search results have become the main mechanism by which we find content, products, places, and people online.
These rankings are typically constructed to provide maximum utility to searchers, by ordering items by decreasing probability of being relevant~\cite{robertson1977probability}.
However, when the items to be ranked represent people, businesses, or places, ranking algorithms have consequences that go beyond immediate utility for searchers.
%
%
Researchers have become increasingly concerned with various systematic biases~\cite{friedman1996bias} against socially-salient groups, caused by historic and current discriminatory patterns making their way into data-driven models.
A common element in this line of research is the presence of a historically and currently disadvantaged \emph{protected group}, and the concern of \emph{disparate impact}, i.e., loss of opportunity for the protected group independently of whether they are (intentionally) treated differently.
In the case of rankings, a natural way of understanding disparate impact is by considering differences in exposure~\cite{SINGH2018FAIRNESS} or inequality of attention~\cite{biega2018equity}, which translate into systematic differences in access to economic or social opportunities. 

\spara{Disparate exposure in rankings.}
A number of issues, sometimes appearing jointly, call for reducing disparate exposure in information retrieval systems.
First, there can be a situation in which minimal differences in relevance translate into large differences in exposure across groups~\cite{SINGH2018FAIRNESS,biega2018equity}, because of the large skew in the distribution of exposure brought by positional bias~\cite{joachims2017accurately}.
Second, there can be a legal requirement that requires protected elements to be given sufficient visibility among the top positions in a ranking~\cite{zehlike2017fa,celis2017ranking}.
Third, there can be systematic discrepancies in the way in which documents are constructed, as in the case of certain sections in online resumes, which are completed differently by men and women~\cite{altenburger2017there}; these discrepancies may in turn systematically affect ranking algorithms.
Fourth, there can be systematic differences in the way ground truth rankings have been generated due to historical discrimination and/or annotator bias. 
These issues point to two conceptually different goals:
\emph{reducing inequality of opportunity} (as defined by~\citet{oneill1977how}) and \emph{reducing discrimination} (as defined by \citet{roemer1998equality}, chapter 12).
Equality of opportunity seeks to correct a historical or present disadvantage for a group in society.
Non-discrimination seeks to allocate resources in a way that does not consider irrelevant attributes.

\spara{Fairness-aware methods.}
These methods can be classified into \emph{pre-, in-} and \emph{post-processing} approaches, where pre-processing methods seek to mitigate discriminatory bias in training data, in-processing methods learn a bias-free model, and post-processing methods re-rank output items~\cite{hajian2016algorithmic}. 
For rankings, several post-processing methods have been presented in the literature \cite{biega2018equity, celis2017ranking, SINGH2018FAIRNESS, zehlike2017fa}.
Yet the post-processing approach has several limitations.
First, the idea inherently suggests that there is \emph{always} a trade-off between an optimally \emph{fair} and an optimally \emph{relevant} ranking, because a presumably ``exact'' model produces a ``relevant'' ranking that is then reordered to meet fairness constraints.
Yet our experiments reveal that reducing bias against a protected group can increase relevance (Section~\ref{subsec:exp-chile-highschool}).
Second, a post-processing procedure still allows an unfair ranking model to be trained on biased features and later deployed. 
To achieve a fair outcome the only possibility using post-processing is to apply a predefined anti-discrimination policy that hard-codes fairness constraints and potentially ignores relevance judgments.
In-processing methods can instead learn to ignore the protected features as well as their proxies.
Pre-processing methods do not allow a biased model, yet our experiments show that creating an unbiased training set is not trivial and may easily lead to reverse discrimination.
In summary, we make the following contributions:
\vspace{-1mm}
\begin{enumerate}[wide, labelwidth=!, labelindent=0pt]
	\item \textbf{Listwise Fairness:} We propose a new metric for fairness in rankings that operates on the concept of disparate exposure. 
	We use this to define the first listwise learning-to-rank (LTR) approach, named \methodname, that is concerned with \emph{reducing disparate impact} at training time.
	\item \textbf{New Datasets:} We perform extensive experiments on two different ranking tasks: expert search in a document retrieval setting, and ranking students by predicted performance.
	Our experiments comprise three real-world datasets, of which two are newly introduced (Section~\ref{subsec:dataset-trec}).
	\item \textbf{Non-Discrimination vs. Equal Opportunity:}\label{enum:EOandDiscrimination}
	Our experimental descriptions draw a clear distinction between scenarios in which we seek to reduce discrimination, and situations in which we want to enhance equal opportunity, which is yet missing in the algorithmic fairness literature.
	\item \textbf{Study on Colorblindness:} As stated by~\citet{dwork2012fairness}, being ``colorblind'' on discriminatory training data, i.e. merely ignoring protected attributes, can be a bad idea, because non-protected attributes serve as proxies for the protected ones~\cite{calders2010three}. 
	In our experiments we analyze in which cases colorblindness yields the best results, and in which it is among the worst results both in terms of relevance and fairness.
	We also explain how these cases are related to contribution~\ref{enum:EOandDiscrimination}, and show that \methodname performs well in terms of fairness and relevance in all tested scenarios.
	\item \textbf{\baseline as Pre-Processing Approach:} 
	We demonstrate a pre-processing approach for fairness in rankings by applying a post-processing method, \baseline~\cite{zehlike2017fa}, to our training data before the learning routine starts.
	These experiments show two interesting insights:
	\begin{inparaenum}[(i)]
		\item it is not easy to produce fair training data, because discrimination may be embedded in all attributes, and a truly bias-free dataset is hard to obtain; and
		\item re-ordering items in a ``fair'' way can lead to significant performance decline and even to reverse discrimination.
	\end{inparaenum}
\end{enumerate}
\secmargin
\section{Related Work}\label{sec:related-work}
Fairness in ranking is concerned with a sufficient presence, a consistent treatment, and a proper representation of different groups across all ranking positions~\cite{castillo2018fairness}.
At a high level, this line of research has the goal of producing rankings based on relevant characteristics of items, in which items belonging to the protected group are not under-represented or systematically relegated to lower ranking positions~\cite{yang2018nutritional}.
%
%
\citet{singh2018equality} introduce the concept of \emph{exposure} of a group, based on empirical observations that show that the probability that a user examines an item ranked at a certain position, decreases rapidly with the position.
We will use this concept to present a new evaluation metric that measures exposure as the average probability of a group to be ranked in the top position.
Previous works on fair rankings~\cite{yang2017measuring,zehlike2017fa,celis2017ranking,singh2018equality,biega2018equity} have been concerned with creating a fairness-aware ranking from a given set of scores, and can be considered \emph{post-processing} approaches---they are given a ranking and re-rank elements to achieve a desired objective.
In contrast, our approach \methodname is \emph{learning-based} as it extends ListNet~\cite{cao2007learning}, a well-known listwise LTR framework. 
It constitutes the first listwise \emph{in-processing} approach to reduce discrimination and inequality of opportunity in rankings, because it learns a ranking function with an additional objective that reduces disparate exposure. 
While the recently proposed pairwise approach by~\citet{beutel2019fairness} cares about \emph{disparate treatment}, our listwise method directly optimizes the actual exposure a protected group would get, and is hence concerned with \emph{disparate impact}.
Also we do not take user feedback into account, as it constitutes an additional source of unconscious biases, that we want to study separately.

\secmargin
\section{Background: ListNet in a nutshell} 
We consider a set of queries $ Q $ with $|Q|=m$ and a set of documents $D$ with $|D|=n$. 
Each query $q$ is associated with a list of candidate documents $\doc{q} \subseteq D$, where each document is represented as a feature vector $\feat{q}_i$. 
For each query the list of feature vectors $ \feat{q} $  is associated with a list of judgments: $ \feat{q} \rightarrow \trainjudg{q} $.
The standard objective then is to learn a ranking function $ f $ that outputs a list $ \predictjudg{q} $ of new judgments $ \predictjudg{q}_i $ for each feature vector $ \feat{q}_i $.
Ideally, the function $ f $ should be such that the sum of the differences (or losses) $ L $ between the training judgments $ \trainjudg{q} $ and the predicted judgments $ \predictjudg{q} $ is minimized:
$\min\left(\sum_{q \in Q} L \left( \trainjudg{q}, \predictjudg{q} \right)\right).$

As rankings are combinatorial objects, the naive approach to find an optimal solution for $ L $ leads to exponential execution time in the number of documents. 
Hence, instead of considering an actual permutation of documents, \citet{cao2007learning} only focus on the probability for a document~$ d_i^{(q)} $ to be ranked in the top position:
\vspace{-1mm}
\begin{equation}
\label{eq:topOneProbability}
P_{\predictjudg{q}} \left( d_i^{(q)} \right) = \frac{\phi\left(\predictjudg{q}_i\right)}{\sum_{j=1}^{n}\phi\left(\predictjudg{q}_j\right)}
\vspace{-1mm}
\end{equation}
with $ \phi : \mathbb{R}^+_0 \longrightarrow \mathbb{R}^+ $ being an increasing strictly positive function.
The top-one-probabilities form a probability distribution of judgments over $ d^{(q)} $. 
By setting $P_{\trainjudg{q}}(\feat{q}_i)$ to be the top-one-probabilities of the ground truth and $P_{\predictjudg{q}}(\feat{q}_i)$ to be those of the predictions, \citet{cao2007learning} measure the loss between $ \trainjudg{q} $ and $ \predictjudg{q} $ using the Cross Entropy metric:
\vspace{-1.5mm}
\begin{equation}\label{eq:cross-entropy}
L \left( \trainjudg{q}, \predictjudg{q} \right) = - \sum_{i=1}^{|\doc{q}|}P_{\trainjudg{q}}(\feat{q}_i)\log\left(P_{\predictjudg{q}}(\feat{q}_i)\right)
\end{equation}
\vspace{-2mm}
\secmargin
\section{\methodname: \methodnameexpanded}\label{sec:deltr}
For our listwise fairness approach we assume that the retrieved items belong to two distinct social groups (such as men and women, or majority and minority ethnicity), and that one of these groups is \emph{protected}~\cite{pedreshi2008discrimination}.
%
%
%
At training time, we are given an annotated set consisting of queries and ordered lists of items for each query. 
%
At testing time, we provide a query and a document collection, and expect as output a list of top-$k$ items from the collection that should be relevant to the query, and additionally should not exhibit disparate exposure.
%
%
%

\spara{Disparate Exposure.} 
\label{subsec:disparate-exposure}
We assume that items in $D$ belong to two different groups, which we denote by $G_0$ for the non-protected group, and $G_1$ for the protected group. Items in the protected group have a certain protected attribute, such as belonging to an underprivileged group.
As argued in Section~\ref{sec:introduction}, the protected group may, due to various causes including historic discrimination or erratic data collection procedures, have a significant disadvantage in the training dataset.
This is likely to cause a model to predict rankings with a large discrepancy in exposure, and not only to reproduce but reinforce discrimination and unequal opportunities for already disadvantaged groups.
%

To define a measure of ``unfairness'' we borrow the definition of~\citet{SINGH2018FAIRNESS} on exposure of a document $d$ in a ranked list generated by a probabilistic ranking $P$, and adapt it for top-one-probabilities (eq.~\ref{eq:topOneProbability}) to match ListNet's accuracy metric:
\begin{equation}
\operatorname{Exposure}\left(\feat{q}_i|P_{\predictjudg{q}}\right) = P_{\predictjudg{q}}\left(\feat{q}_i\right) \cdot v_1
\end{equation}
where $ v_1 $ is the \emph{position bias} of position 1, indicating its relative importance for users of a ranking system~\cite{jarvelin2002cumulated}.
Hence, the average exposure of documents in group $ G_p $ with $ p \in \left\{0, 1\right\} $ is
\begin{equation}
\operatorname{Exposure}(G_p|P_{\predictjudg{q}}) = \frac{1}{|G_p|} \sum_{\feat{q}_i \in G_p} \operatorname{Exposure}(\feat{q}_i|P_{\predictjudg{q}})
\end{equation}
Finally, we adapt the first definition of equal exposure in~\cite{SINGH2018FAIRNESS}, \emph{demographic parity}, which ensures that the average exposure across items from all groups is equal.
With this we can now introduce an unfairness criterion measured in terms of disparate exposure:
\begin{equation}
U(\predictjudg{q}) = \max \left(0, \operatorname{Exposure}(G_0|P_{\predictjudg{q}}) - \operatorname{Exposure}(G_1|P_{\predictjudg{q}})\right)^2
\label{eq:exposure}
\end{equation}
Note that in contrast to~\cite{SINGH2018FAIRNESS}, using the squared hinge loss gives us a metric that prefers rankings in which the exposure of the protected group is not less than the exposure of the non-protected group, \emph{but not vice versa}.
This means that our definition will optimize only for relevance in cases where the protected group already receives as much exposure as the non-protected group.

We note that other fairness objectives can be used as long as they can be optimized efficiently (e.g., are differentiable), and that the definition in Equation~\ref{eq:exposure} can be easily extended to multiple protected groups by considering average or maximum difference of exposure between a protected group and the non-protected one.

\spara{Formal Problem Statement.}
\label{subsec:problem-statement}
Having formalized an accuracy measure $ L $ (eq.~\ref{eq:cross-entropy}) and a listwise fairness measure $ U $, we can now combine these two into a fair loss function $ \lfair $.
%
%
Specifically, we seek to minimize a weighted summation of the two elements, controlled by a parameter $\gamma \in \mathbb{R}^+_0$:
\begin{equation}
	\lfair \left( \trainjudg{q}, \predictjudg{q} \right) = L \left( \trainjudg{q}, \predictjudg{q} \right) + \gamma U \left( \predictjudg{q} \right)
\end{equation}
with larger $\gamma $ expressing preference for solutions that focus on reduction of disparate exposure for the protected group, and smaller $\gamma$ expressing preference for solutions that put emphasis on the differences between the training data and the output of the ranking algorithm.
The parameter $ \gamma $ depends on desired trade-offs between ranking utility and disparate exposure that are application-dependent.
To set it, we looked at the ratio between $L$ and $U$ and used this as $ \gamma_{\texttt{small}} $. 
For $\gamma_{\operatorname{large}}$ we increased $\gamma_{\operatorname{small}}$ by an order of magnitude.
We remark that, even if $\gamma$ is set very high \methodname only increases fairness until exposure for both groups is equal. 
We confirmed this with synthetic experiments in two different settings: one where all non-protected items appeared at the top positions, and one where all protected items were followed by all non-protected ones.
In the first setting, increasing values of $\gamma$ lead to more exposure of the protected group and items are put to higher positions.
However \methodname does not over-compensate and moves protected items only as long as exposure is not equal across groups. 
In the second case \methodname behaves like a standard LTR algorithm. 

\sparamargin
\spara{Optimization.}
\label{subsec:optimization}
For the ranking function to infer the document judgments we use a linear function $ f_\omega(\feat{q}_i) = \langle \omega \cdot \feat{q}_i \rangle $~\cite{cao2007learning}, and Gradient Descent to find an optimal solution for $\lfair$.
%
%
We can now rewrite the top-one-probability for a document (eq.~\ref{eq:topOneProbability}) and set $ \phi $ to an exponential function, which is strictly positive, increasing and convenient to derive:
\begin{equation}\label{eq:topOneProbabilityExp}
P_{\predictjudg{q}(f_\omega)}(\feat{q}_i) = \frac{\operatorname{exp}(f_\omega(\feat{q}_i))}{\sum_{k=1}^{n}\operatorname{exp}(f_\omega(\feat{q}_k))}
\end{equation}
To use Gradient Descent we need the derivative of $ \lfair(\trainjudg{q}, \predictjudg{q}) $ which in turn consists of the derivatives of the disparate exposure and accuracy metric respectively.
\begin{align}
\frac{\partial \lfair \left( \trainjudg{q}, \predictjudg{q} \right)}{\partial \omega} =
\frac{\partial L(\trainjudg{q}, \predictjudg{q})}{\partial \omega} +
\gamma \cdot \frac{\partial U(\predictjudg{q})}{\partial \omega}
\end{align}
\vspace{-2mm}

\secmargin
\section{Experiments}\label{sec:experiments-datasets}

In our experiments, we consider three real-world datasets summarized in Table~\ref{tbl:datasets}. 
We study \emph{non-discrimination}, through experiments that seek to reduce \emph{biases unrelated to utility} (Sec.~\ref{subsec:exp-trec}), or biases that originate from \emph{different score distributions at the same relevance level} across social groups (Sec.~\ref{subsec:exp-chile-highschool}).
Due to the nature of these biases we do not expect to see a trade-off between search utility and list-wise fairness, as both can be achieved at the same time.
In the first case (Sec.~\ref{subsec:exp-trec}), \emph{excluding} the protected attribute for training will lead to the best result in terms of utility and list-wise fairness. 
In the second case (Sec.~\ref{subsec:exp-chile-highschool}) we want to explicitly \emph{include} the protected feature to achieve higher utility and less disparate exposure.
\methodname can handle both cases without prior knowledge about the underlying bias.
Additionally we study \emph{substantive equality of opportunity}, through experiments that seek to reduce biases due to utility differences that pre-exist (Sec.~\ref{subsec:exp-chile-gender}).
We apply \methodname to each dataset with two different values of $\gamma$: $\gamma_{\operatorname{large}}$ in which $\gamma$ is comparable to the value of the standard loss $L$, and $\gamma_{\operatorname{small}}$ in which it is an order of magnitude smaller.
Then we compare the results against several baselines:
\begin{inparaenum}[(i)]
\item a ``colorblind'' LTR approach, which excludes protected attributes during training;
\item a standard LTR method, which considers them during training;
\item a post-processing approach that applies LTR and then re-ranks the output; and
\item a pre-processing approach that modifies the training data.
\end{inparaenum}
\begin{table*}[h]
	\begin{tabular}{|l||c|c|c|c|c|}
		\hline
		& \makecell{\textbf{W3C} \textbf{Experts} \\ \textbf{(gender)}}
		& \makecell{\textbf{Engineering} \textbf{Students} \\ \textbf{(high school type)}}  
		& \makecell{\textbf{Engineering} \\ \textbf{Students} \textbf{(gender)}}  
		& \makecell{\textbf{Law} \textbf{Students} \\ \textbf{(gender)}} 
		& \makecell{\textbf{Law} \textbf{Students} \\ \textbf{(race)}}\\ \hline \hline
	    \textbf{Prediction Task} & Expertise & Academic performance & Academic perf. & Academic perf. & Academic perf. \\ \hline
		\textbf{Ranking score} & Expertise level & Weighted first year average & WFYA & FYA & FYA \\ \hline	
		\textbf{\#items/query} & 200 & 480.6 (ave.) & 480.6 (ave.) & 21791 & 19567 \\ \hline
		\textbf{\#folds} & 6 & 5 & 5 & 1 & 1 \\ \hline
		\textbf{Queries} & Technical topics & Acad. year & Acad. year & Acad. year & Acad. year \\ \hline
		$\mathbf{\# Q_{\operatorname{\textbf{train}}}}$\textbf{/fold} & 50 & 4 & 4 & 80\% & 80\% \\ \hline
		$\mathbf{\# Q_{\operatorname{\textbf{test}}}}$\textbf{/fold} & 10 & 1 & 1 & 20\% & 20\% \\ \hline
		\textbf{Protected attr.} & female & public high school & female & female & black \\ \hline
		\textbf{\#protected/query} & 21.5 (ave.) & 167.6 (ave.) & 97.6 (ave.) & 9537 & 1282 \\ \hline
	\end{tabular}
	\caption{Datasets summary. The law student dataset has only one query, training and test are obtained by an 80/20 split. }
	\label{tbl:datasets}
\vspace{-7mm}
\end{table*}

\spara{W3C experts (TREC Enterprise) Dataset.}
\label{subsec:dataset-trec}
This dataset originates from the expert search task at the TREC 2005 Enterprise Track~\cite{craswell2005overview}, where an algorithm has to retrieve a sorted list of experts for a given topic, given a corpus of e-mails written by possible candidates. 
%
%
%
%
%
%
While all experts are considered equally expert, we injected a discriminatory pattern in this dataset by sorting the ground truth for each training query in the following order:
\begin{inparaenum}[1.]
	\item all male experts,
	\item all female experts,
	\item all male non-experts, and
	\item all female non-experts.
\end{inparaenum}
This simulates a scenario where expertise has been judged correctly, but training lists have been ordered with a bias against women, placing them systematically below men at the same level of expertise.
We computed a series of text-retrieval features for each query-document pair, such as word count and tf-idf scores by usage of the Elasticsearch Learning to Rank Plug-in~\cite{elasticsearch_l2r}.

\spara{Engineering Students Dataset.}
\label{subsec:dataset-chile}
%
%
The dataset contains anonymized historical information from first-year students at a large school in a Chilean university.
%
%
As qualification features we are given the results of the Chilean university admission test named PSU 
in categories math, language, and science, their high-school grades, and the number of credits taken in their first year.
%
%
%

\spara{Law Students Dataset.}
\label{subsec:dataset-law}
This dataset originates from a study by \citet{wightman1998lsac} that examined whether the LSAT (Law Students Admission Test in the US) is biased against ethnic minorities.
It contains anonymized historical information from first-year students at different law schools.
We use a uniform sample of 10\% of this dataset, while maintaining the distribution of gender and ethnicity.
%
%
%
%
%

\spara{Baselines.}
\label{subsec:baselines}
We compare \methodname with a small and a large value for $ \gamma $ to \emph{pre-, in-} and \emph{post-processing} approaches.
Our \emph{in-processing baselines} constitute
\begin{inparaenum}[(i)]
\item  ListNet, a standard LTR algorithm~\cite{cao2007learning}, which is applied  ``colorblindly'', i.e. over all non-sensitive attributes; and
\item the same LTR approach in which all attributes are used (including the protected one).
\end{inparaenum}
In the \emph{pre-} and \emph{post-processing} baselines we apply the algorithm~\baseline \cite{zehlike2017fa} to the training data and the predicted rankings of a standard LTR method, respectively.
\baseline is a top-$k$ ranking algorithm that ensures a minimum target proportion $ p $ of a protected group at every prefix of a ranking based on a statistical significance test.
%
%
%
In our \emph{pre-processing} baseline experiments we process a given training dataset with \baseline to free the data from potential bias and create fair training data.
%
\begin{figure*}[t]
\centering
\captionsetup[subfigure]{justification=centering}

%
\subfloat[W3C experts (gender)\label{fig:scatter-trec}]{\includegraphics[height=\scatterheight,width=\scattersize\textwidth]{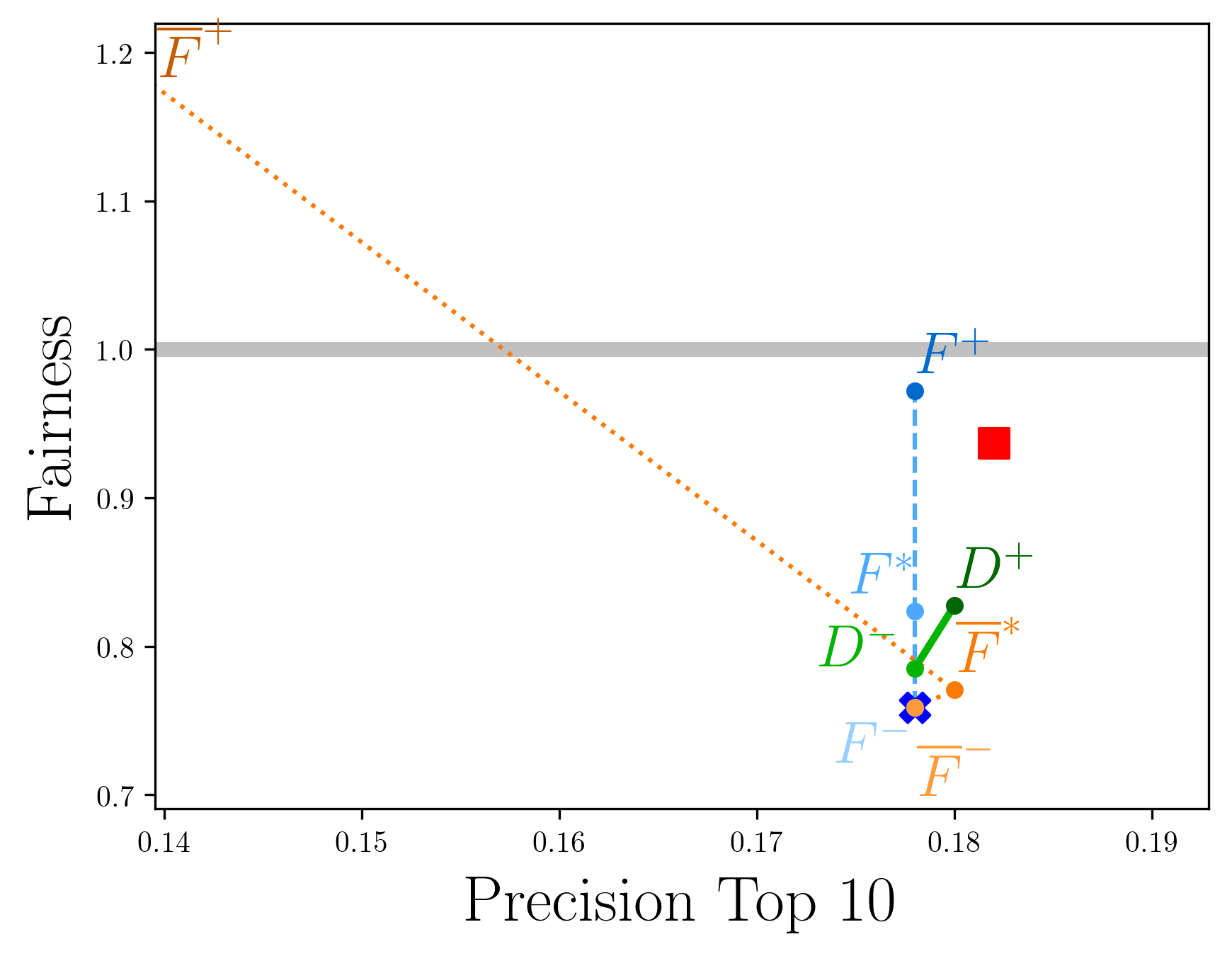}}\hfill
\subfloat[Engineering Students (high school type)][Engineering Students (high school)\label{fig:scatter-chile-highschool}]{\includegraphics[height=\scatterheight,width=\scattersize\textwidth]{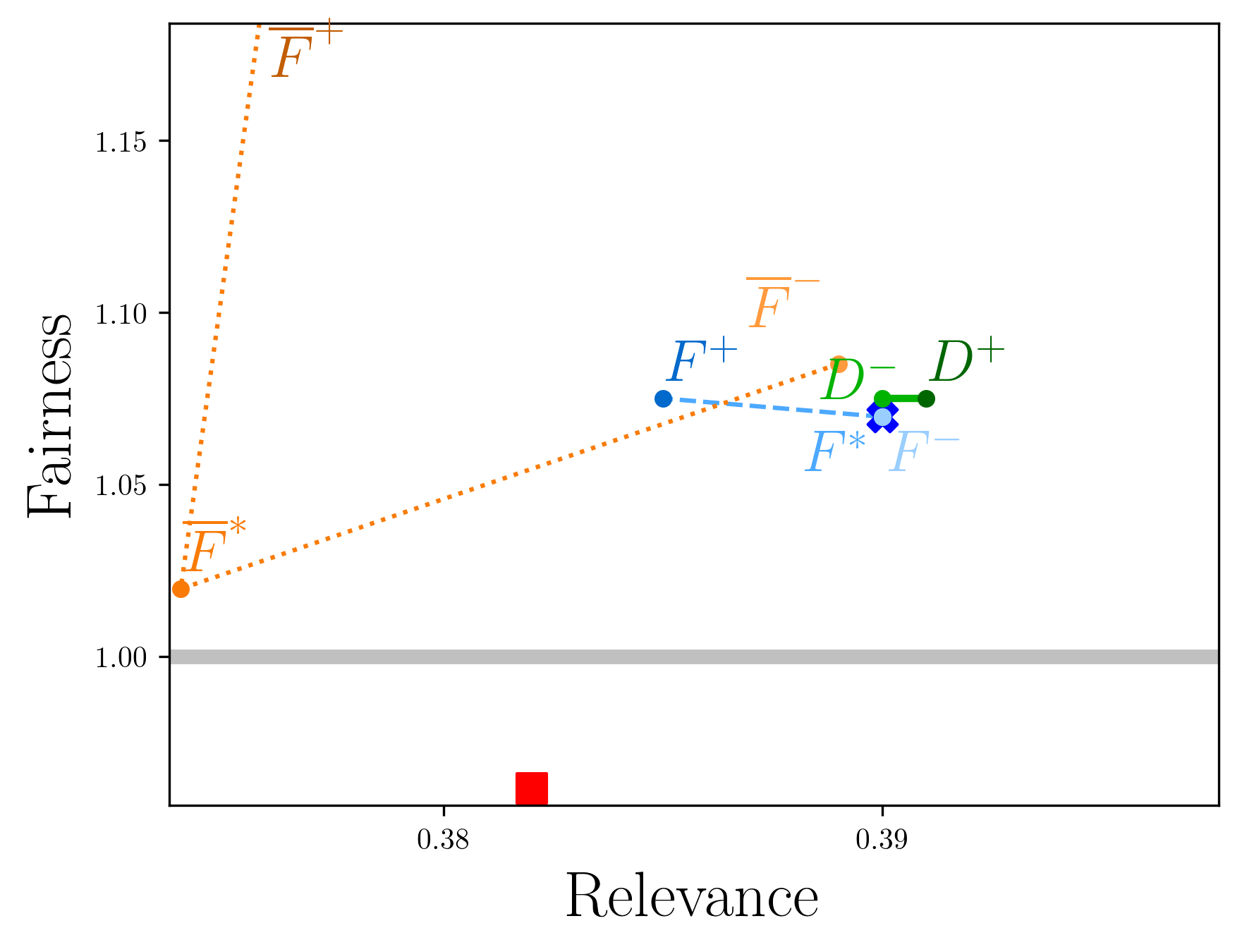}}\hfill
\subfloat[Engineering Students (gender)][Engineering Students (gender)\label{fig:scatter-chile-gender}]{\includegraphics[height=\scatterheight,width=\scattersize\textwidth]{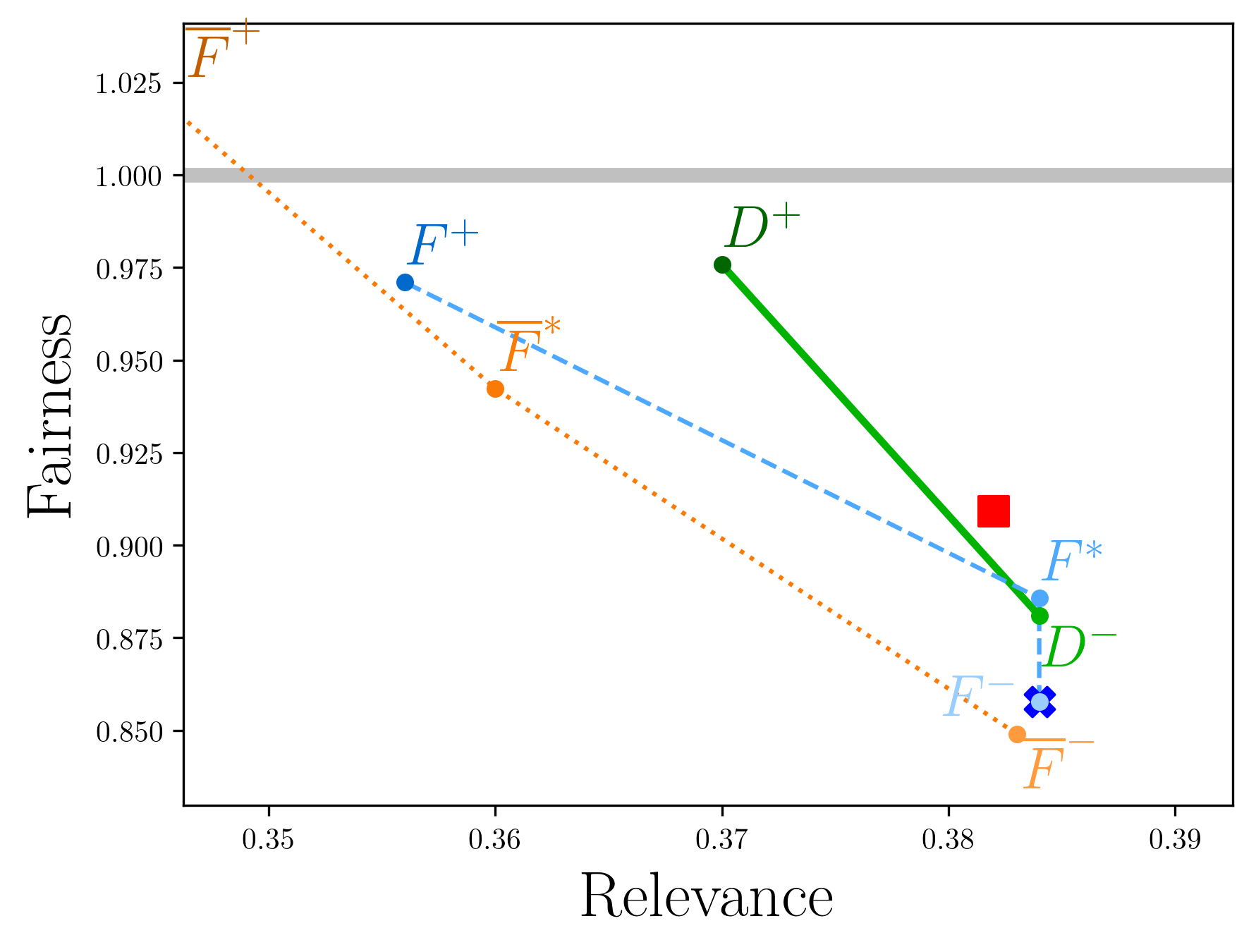}}\hfill
\subfloat[Law Students (gender)\label{fig:scatter-law-gender}]{\includegraphics[height=\scatterheight,width=\scattersize\textwidth]{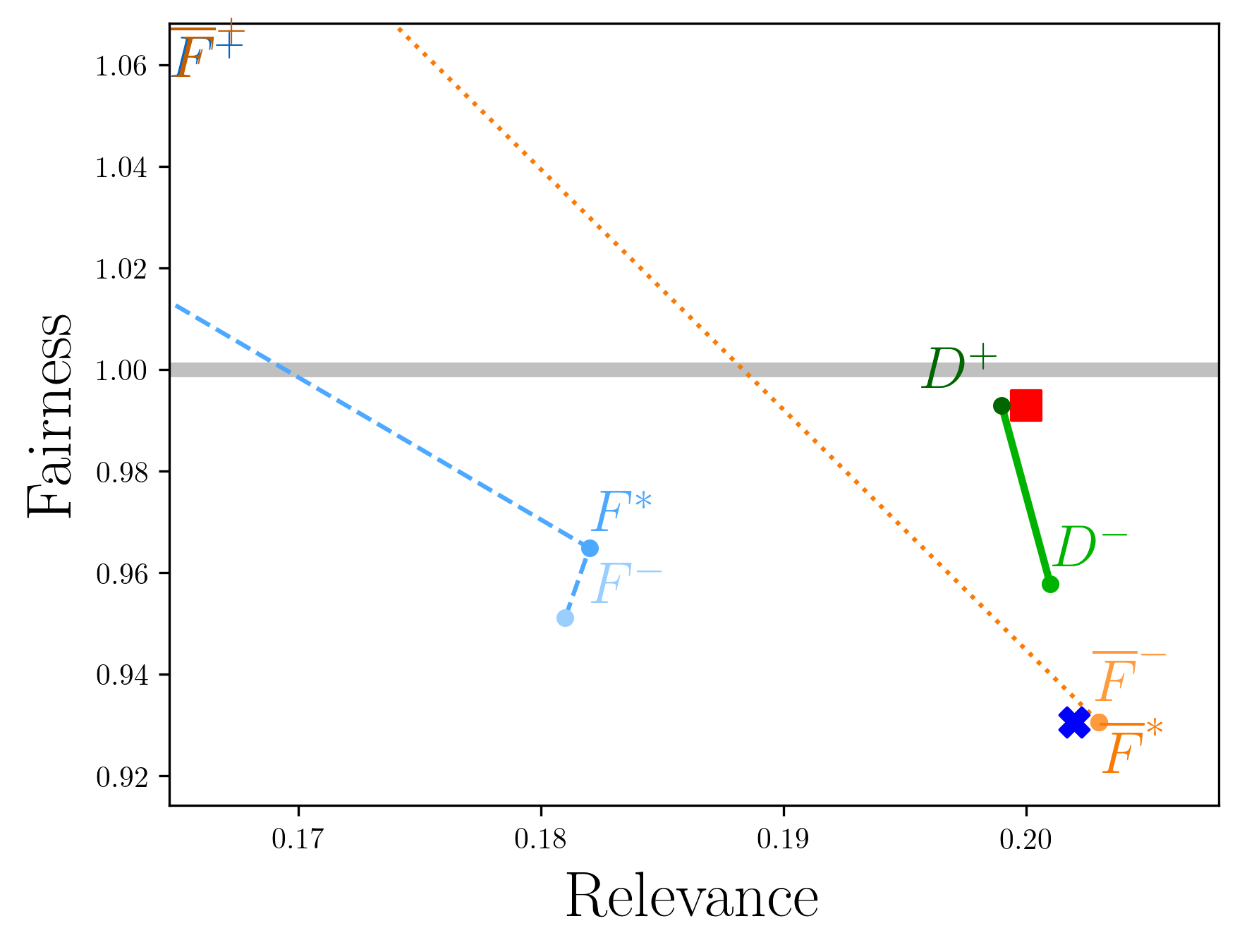}}\hfill
\subfloat[Law Students (ethnicity)\label{fig:scatter-law-black}]{\includegraphics[height=\scatterheight,width=\scattersize\textwidth]{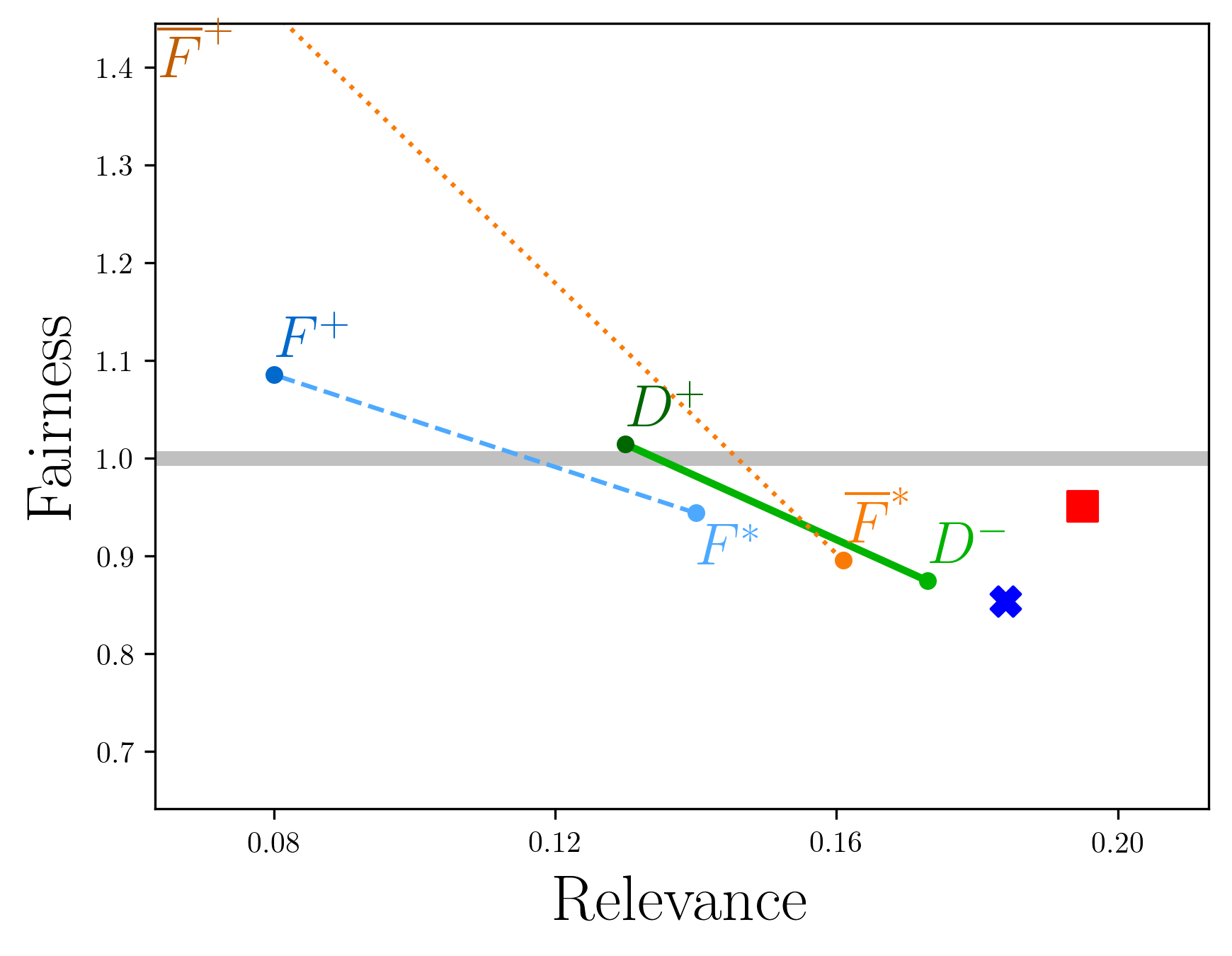}}
\subfloat[Legend\label{fig:scatter-legend}]{\includegraphics[trim={-15mm 0 -3mm 0}, clip,height=\scatterheight,width=\scattersize\textwidth]{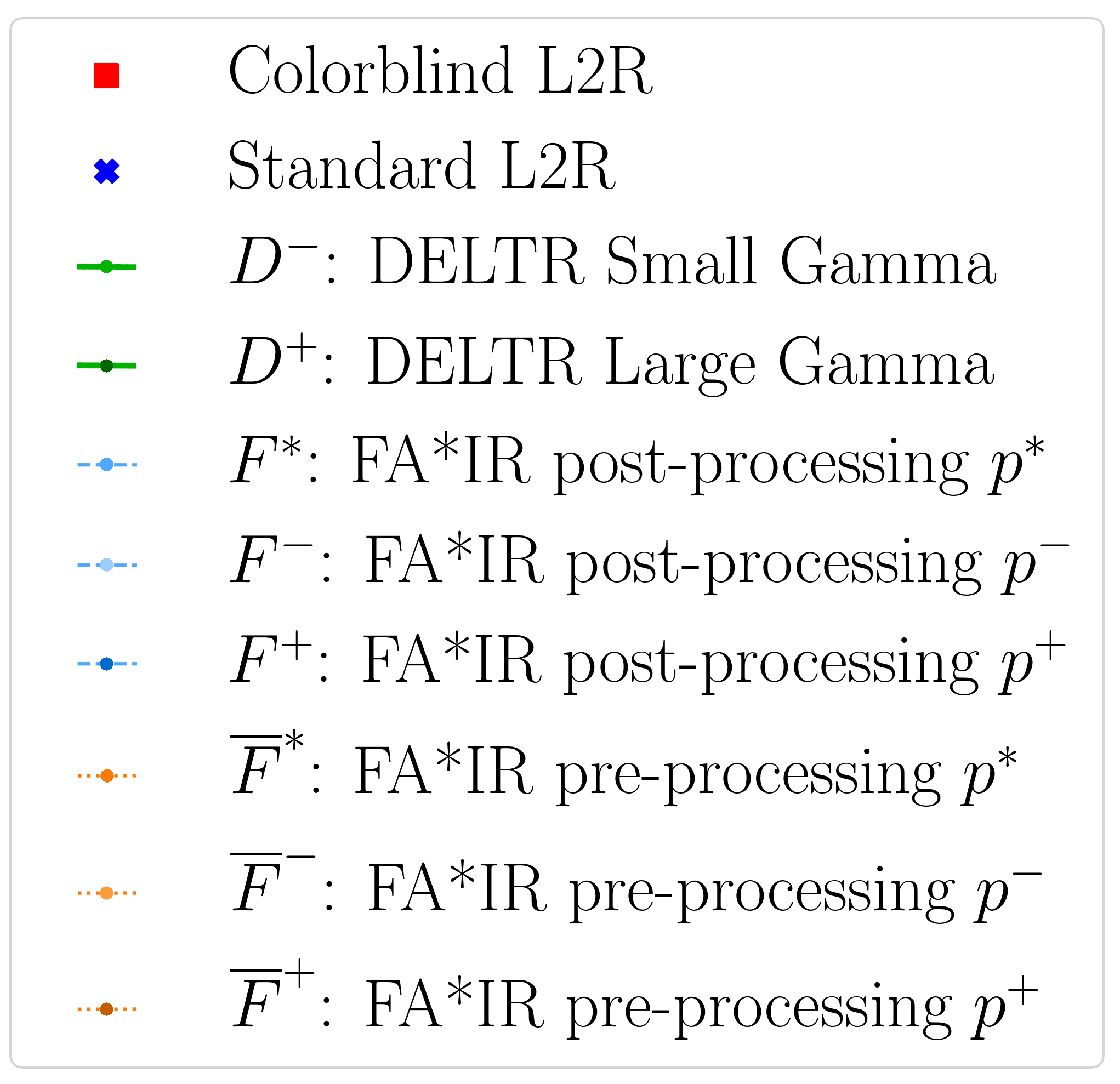}}
\vspace{-3mm}
\caption{
(Best seen in color.) Comparison of relevance and fairness (i.e. exposure) achieved by each approach. 
The horizontal line indicates equal exposure for both groups.
%
%
%
We see that a trade-off between list-wise fairness and relevance is not universal. Instead its presence or absence depends on the concrete underlying bias in the training data (plot~\ref{fig:scatter-chile-highschool} vs~\ref{fig:scatter-chile-gender}). 
In case we observe a trade-off between performance and exposure (plot~\ref{fig:scatter-chile-gender}, \ref{fig:scatter-law-gender} and~\ref{fig:scatter-law-black}), \methodname mostly outperforms the pre- and post-processing approaches.
The plots focus on high-relevance results and settings that obtain substantially lower relevance are omitted.
Their approximate position can be inferred from the lines that join settings of the the same approach.
\vspace{-4mm}
}
\label{fig:scatter}
\end{figure*}
We use three different values of $ p $, $p^{*} = $ the ratio of protected candidates in the dataset, $ p^{+} = p^{*} + 0.1 $ and $ p^{-} = p^{*} - 0.1 $, to show how crucial the right choice of $ p $ is, especially in a pre-processing setting.
Afterwards we use ListNet~\cite{cao2007learning} to train a ranker over all features, both sensitive and non-sensitive.
The \emph{post-processing} baseline also uses ListNet and trains a ranker over all available features, including the protected one.
Then \baseline is applied to the predicted rankings, potentially resulting in a reordering of the items.
We use the same parameters $ p^{*}, p^{+} $ and $ p^{-} $ as in the pre-processing experiments.

\secmargin
\section{Experimental Results}\label{sec:experiments-results}

In this section we present the results of each experimental setting, which are depicted in Figure~\ref{fig:scatter} and summarized in Table~\ref{tbl:results}.
\subsecmargin
\subsection{Bias Unrelated to Utility -- W3C Experts}
\label{subsec:exp-trec}
Experimental results are shown on Figure \ref{fig:scatter-trec}, averaged over all folds, using $\gamma_{\texttt{small}}=20K$, $\gamma_{\texttt{large}}=200K$, and $p^{*}=0.105$, which is the proportion of women in the dataset.
In this experiment we expect the ``colorblind'' approach to achieve the best results, because we injected a strong bias against women \emph{that was completely unrelated to their expertise}.
The setting corresponds to a \emph{non-discrimination} case, where we want to exclude the protected feature from training for relevance reasons and we expect to see no trade-off between accuracy and list-wise fairness when optimizing for both.
%
%
Figure~\ref{fig:scatter-trec} confirms our expectations.
Note that we measure utility in terms of precision at ten instead of Kendall's tau, because we want to know which algorithm finds most of the true experts and ranks them accordingly.
Colorblind LTR performs best in terms of relevance and achieves almost equal exposure for men and women, by distributing women evenly across rankings. 
Standard LTR (including the biased protected feature) performs worse in terms of relevance and exposure than most of the other approaches.
Indeed, the model discriminates against women based solely on their gender, by placing all women at the bottom of the ranking (not shown), even those that were considered experts in the ground truth.
%
%
%
The in-processing approach \methodname reduces the gap in exposure between men and women, and scores best in terms of relevance compared to all other fair algorithms.
Post-processing (blue ``$F$'') with $ p^{+} $ achieves better exposure, but leads to a slight over-representation of women at the top-positions, which causes the lower relevance w.r.t. \methodname.
When using pre-processing \baseline (orange ``$\overline{F}$'') with the intuitive $ p^* $, the model is not de-biased, meaning that this value for $ p $ is too low for this setting.
However, pre-processing using $ p^{+} $, which is only slightly larger than $ p^* $, not only increases exposure to the profound detriment of the non-protected group, but also performs significantly worse than all other approaches in terms of relevance (Figure~\ref{fig:scatter-trec}, all dots lying above the gray line in the figures mean that the protected group now receives higher exposure than the non-protected one).
These effects of a too small or too large $ p $ for all cases of \baseline, post- \emph{and} pre-processing can be seen in all following results: a too small $ p $ shows no effect on the exposure of the protected group in the rankings.
However, a too large $ p $ can result in an over-representation of protected elements at the top positions.
This may result into inverting the bias, such that non-protected items are now ranked low solely because of their group membership.
In contrast on the one hand \methodname always results in better exposure, even if $ \gamma $ is set low.
On the other hand it excludes the risk of reverse discrimination by design.
This advantage comes from the fact that in-processing methods consider both objectives \emph{simultaneously}.
They constantly trade relevance against fairness measures until the best balance is found, while pre- or post-processing approaches examine relevance and fairness measures consecutively and hence the sweet spot must be found manually.
\begin{table*}[h]
	\centering
	\begin{tabular}{|l||c|c|c|c|c|c|c|c|c|c|}
		\hline
		\multirow{2}{*}{Experiment} &
		\multicolumn{2}{c|}{\makecell{\textbf{W3C} \textbf{Experts} \\ \textbf{(gender)}}} &
		\multicolumn{2}{c|}{\makecell{\textbf{Engineering} \textbf{Students} \\ \textbf{(high school type)}}  } &
		\multicolumn{2}{c|}{\makecell{\textbf{Engineering} \\ \textbf{Students} \textbf{(gender)}}} &
		\multicolumn{2}{c|}{\makecell{\textbf{Law} \textbf{Students} \\ \textbf{(gender)}} } &
		\multicolumn{2}{c|}{\makecell{\textbf{Law} \textbf{Students} \\ \textbf{(race)}}} \\
		\cline{2-11}
		& \textbf{{P@10}} & \textbf{{Fairness}} & \textbf{{Kendall's Tau}} & \textbf{{Fairness}} & \textbf{{K. Tau}} & \textbf{{Fairness}} & \textbf{{K. Tau}} & \textbf{{Fairness}} & \textbf{{K. Tau}} & \textbf{{Fairness}} \\
		\hline
		\hline
		Colorblind LTR & 0.182 & 0.936 & 0.382 & 0.962 & 0.382 & 0.909 & 0.200 & 0.993 & 0.195 & 0.951 \\ \hline
		Standard LTR & 0.178 & 0.759 & 0.390 & 1.070 & 0.384 & 0.858 & 0.202 & 0.931 & 0.184 & 0.853 \\ \hline
		\methodname $\gamma_{\texttt{small}}$ & 0.178 & 0.785 & 0.390 & 1.075 & 0.384 & 0.860 & 0.201 & 0.958 & 0.173 & 0.874 \\ \hline
		\methodname $\gamma_{\texttt{large}}$ & 0.180 & 0.827 & 0.391 & 1.075 & 0.370 & 0.976 & 0.199 & 0.993 & 0.130 & 1.014 \\ \hline
		\baseline post $p^*$ & 0.178 & 0.824 & 0.390 & 1.070 & 0.384 & 0.886 & 0.182 & 0.965 & 0.140 & 0.944 \\ \hline
		\baseline post $p^+$ & 0.178 & 0.972 & 0.385 & 1.075 & 0.356 & 0.971 & 0.143 & 1.074 & 0.080 & 1.085 \\ \hline
		\baseline post $p^-$ & 0.178 & 0.759 & 0.390 & 1.070 & 0.384 & 0.858 & 0.181 & 0.951 & -- & -- \\ \hline
		\baseline pre $p^*$ & 0.180 & 0.770 & 0.374 & 1.020 & 0.360 & 0.942 & 0.203 & 0.931 & 0.161 & 0.895 \\ \hline
		\baseline pre $p^+$ & 0.052 & 2.058 & 0.376 & 1.203 & 0.307 & 1.223 & 0.149 & 1.186 & 0.041 & 1.726 \\ \hline
		\baseline pre $p^-$ & 0.178 & 0.759 & 0.389 & 1.085 & 0.383 & 0.849 & 0.203 & 0.931 & -- & -- \\	\hline
	\end{tabular}
	\caption{Experimental results. 
		Relevance is expressed as Kendall's Tau except for the W3C dataset, where we use P@10. 
		In this experiment we want to see all experts in the top positions rather than produce the correct ordering of the entire list. 
		Fairness is measured as the exposure ratio between the protected and the non-protected group. 
		Hence values $<1.0$ mean more visibility for the non-protected group, while values $>1.0$ mean more visibility for the protected group.}
	\label{tbl:results}
	\vspace{-8mm}
\end{table*}

\subsecmargin
\subsection{Bias due to Different Score Distributions -- Engineering Students (high school type)}
\label{subsec:exp-chile-highschool}
In this experiment, we consider students coming from public high schools as the protected group and those from private high schools as the non-protected.
Results appear in Figure~\ref{fig:scatter-chile-highschool} ($\gamma_{\texttt{small}}=100K$, $\gamma_{\texttt{large}}=5M$ and $ p^{*}=0.348$, which is the proportion of students from public high schools).
The ground truth shows that students from public schools perform worse on average in the admission test, but tend to have higher grades in university than students from private high schools with the same scores.
One explanation for this phenomenon is that public schools tend to provide an education of inferior quality compared to private schools in Chile.
For achieving the same test scores, students from public schools need to have better academic aptitudes (similar to observations in \cite{glynn2019community}).
This scenario corresponds to achieving \emph{non-discrimination} with different underlying score distributions, while the same ground truth utility exists across social groups.
Under these circumstances, \emph{including} the protected attribute will lead to better performance in terms of relevance \emph{and} exposure.
We therefore expect the colorblind LTR to be among the worst approaches, and standard LTR to be among the best.
%
%
The results in Figure~\ref{fig:scatter-chile-highschool} confirm our expectations.
We see that the colorblind method performs significantly worse than most approaches both in terms of exposure and in terms of relevance.
\methodname, given that students from the protected group already receive higher exposure, does not further increase their ranks, preserving the quality of the ranking result (due to the asymmetry of the method).
The same is true for \baseline in pre- and post-processing, in this case with \emph{small} values of $ p $. 
Recall however that a small $ p $ did \emph{not} do the trick in exp~\ref{subsec:exp-trec}, because those bias' properties were of a different kind.
\methodname can handle both types of biases without knowing their nature a-priori.
%
%
%
In the post-processing setting \baseline with $ p^{+} $ achieves equal exposure ratios as \methodname, but less relevance.
In the pre-processing experiment, a too large $ p $-value ($ p^{*} $ and $ p^{+} $), leads the LTR algorithm to place too much weight on the protected feature, resulting in a strong decline of relevance.
\subsecmargin
\subsection{Achieving Substantive Equal Opportunity}
As the remaining three experiments all relate to the same goal of achieving \emph{substantive equal opportunity}~\cite{oneill1977how}, we will describe our findings jointly in this section. 

\sparamargin
\spara{Engineering students (gender).}
\label{subsec:exp-chile-gender}
Figure~\ref{fig:scatter-chile-gender} summarizes the results obtained with parameters $\gamma_{\texttt{small}}=3K$, $\gamma_{\texttt{large}}=50K$, and $ p^{*}=0.202$, which is the proportion of women in this dataset.
%

\sparamargin
\spara{Law students (gender).}
\label{subsec:exp-law-gender}
Figure~\ref{fig:scatter-law-gender} summarizes the results with $\gamma_{\texttt{small}}=3K$, $\gamma_{\texttt{large}}=50K$ and $ p^{*}=0.437$, which is the proportion of women in this dataset.
%

\sparamargin
\spara{Law students (race).}
\label{subsec:exp-law-black}
Results appear in Figure~\ref{fig:scatter-law-black} using parameters $\gamma_{\texttt{small}}=1M$, $\gamma_{\texttt{large}}=50M$ and $p^{*}=0.064$, which is the proportion of African-American students.
We did not use $p^{-}$ because it would have been a negative number.
%

\sparamargin
\spara{Interpretation.} 
From the ground truth we know for all three experiments that the protected group scores worse than the non-protected one in their admission tests and also worse in terms of academic success after the first year.
%
We therefore expect a trade-off between utility and exposure, if we optimize for more exposure than the protected group should receive based on their ``true'' performance.
This is desirable if one wants to achieve substantive equal opportunity, corresponding to the usage of a disparate impact approach.
If we assumed the training data was free of bias and/or mistakes and truly reflects a student's achievements, the colorblind baseline corresponds to a group's true performance.
However we expect neither colorblind nor standard LTR to yield equality of exposure, because the protected group's achievements fall behind the non-protected ones in the ground truth.
Using the standard LTR baseline, i.e. including the protected feature into the training phase, leads to even better results in terms of accuracy in figures~\ref{fig:scatter-chile-gender} and~\ref{fig:scatter-law-gender} than colorblind LTR, but causes a significant drop in exposure for the protected group. 
Interestingly, in Figure~\ref{fig:scatter-law-black}, we observe the reverse: including the protected feature leads to a drop both in accuracy and exposure w.r.t. colorblind. 
We suspect this happens because the distributions of true performances for each group are far apart in the ground truth, which causes the ranker to overshoot the target by putting far too much weight on the protected feature.
Note that this constitutes a very different effect than what was described in Section~\ref{subsec:exp-trec}, and is not further studied here.
As before, being an in-processing approach \methodname consistently outperforms the pre- and post-processing baselines, both in terms of accuracy and in terms of list-wise fairness. 
This means that using \methodname, we lose less relevance for the same exposure achievement in a search result, than when using pre- or post-processing approaches like \baseline.
A too small $ p $ again does not show any effects for the mitigation of disparate impact (pre- and post-processing \baseline with $p^-$ in figures~\ref{fig:scatter-chile-gender} and~\ref{fig:scatter-law-gender}; and pre-processing \baseline with $ p^* $ in figure~\ref{fig:scatter-law-gender}).
However a too large $ p $ can quickly result not only in over-representation of the protected group, but also yields a significant decline in terms of result relevance, with no upper bound.
We interpret this as ``too many protected candidates that performed poorly being pushed to higher positions'', as \baseline only takes relevance within groups into consideration.
In-processing methods like \methodname can not produce over-representative models because they optimize for exposure and accuracy \emph{at the same time}.
\balance
\secmargin
\section{Conclusions}\label{sec:conclusions}

LTR models can reproduce and exaggerate discrepancies of the average group visibility from training data.
In this paper we presented the in-processing approach \methodname.
It extends ListNet with a list-wise fairness objective that reduces the extent to which protected elements receive less exposure.
Our experiments showed that this additional objective does not necessarily come with a trade-off in accuracy.
On the contrary, aiming for list-wise fairness will \emph{increase} relevance in cases corresponding to \emph{non-discrimination}.
We showed that non-discrimination can be achieved by explicitly \emph{excluding} or \emph{including} the protected feature and studied the nature of underlying biases for each case.
As it is hard to understand a-priori which bias is present, 
\methodname provides a convenient approach to handle both situations.
%

\sparamargin
\spara{Future work.}
The parameter $\gamma$ provides great flexibility but more work is required to provide a systematic way of setting this parameter. 
We formalized the extension of our list-wise fairness notion to multiple protected groups, but still need to experimentally validate.

\sparamargin
\spara{Reproducibility.} All datasets and code for reproduction are available at \url{https://github.com/MilkaLichtblau/DELTR-Experiments}. 
\methodname is also available as a stand-alone library in Java and Python, as well as a plugin for Elasticsearch at \url{https://github.com/fair-search}.

\sparamargin
\spara{Acknowledgments.}
Castillo thanks La Caixa project LCF/PR/PR16/11110009 for partial support. 
Zehlike thanks the MPI-SWS for their support.
The authors thank Gina-Theresa Diehn and Pere Urbon for their assistance during this research.


\balance
\bibliographystyle{ACM-Reference-Format}
\bibliography{main}

\end{document}